\documentclass[aps,prd,
amsmath,amssymb,nofootinbib,superscriptaddress]{revtex4-2}
\usepackage{graphicx}  
\usepackage{color}
\usepackage{times}
\usepackage[utf8]{inputenc} 
\usepackage{bm}
\usepackage{ulem}
\usepackage{multirow}
\usepackage{makecell}
\usepackage{url}
\usepackage{natbib}
\usepackage{mathrsfs}
\usepackage{physics}
\usepackage{comment}
\usepackage[table,xcdraw]{xcolor}
\usepackage{booktabs,makecell}
\usepackage{amsmath}
\usepackage{pifont}
\usepackage[colorlinks=true,citecolor=blue,urlcolor=blue,linkcolor=blue]{hyperref}
\usepackage{microtype}
\usepackage[none]{hyphenat}

\newcommand{\dalm}{\kern1pt\vbox{\hrule height 0.9pt\hbox{\vrule width 0.9pt
\hskip 2.5pt\vbox{\vskip 5.5pt}\hskip 3pt\vrule width 0.3pt}\hrule height 0.3pt}
\kern1pt}

\begin{document}



\title{Universality of dual mass-scaled fundamental modes in two-fluid neutron stars with mirror dark matter
}


\author{Hajime Sotani}
\email{sotani@yukawa.kyoto-u.ac.jp}
\affiliation{Department of Mathematics and Physics, Kochi University, Kochi, 780-8520, Japan}
\affiliation{RIKEN Center for Interdisciplinary Theoretical and Mathematical Sciences (iTHEMS), RIKEN, Wako 351-0198, Japan}
\affiliation{Kochi University of Technology, Kochi, 780-8515, Japan.}
\affiliation{Theoretical Astrophysics, IAAT, University of T\"{u}bingen, 72076 T\"{u}bingen, Germany}

\author{Ankit Kumar}
\email{ankitlatiyan25@gmail.com}
\affiliation{Department of Physics, Indian Institute of Science, Bangalore 560012, India}
\affiliation{CEICO, Institute of Physics of the Czech Academy of Sciences (FZU), Na Slovance 1999/2, 182 00 Prague 8, Czech Republic}

\date{\today}

\begin{abstract}
Universal relations provide a particularly useful way to extract physical information from neutron star observables in the presence of various uncertainties by reducing the dependence on uncertain model parameters and microphysical inputs. 
In this study, we examine the oscillation frequencies of mirror dark matter admixed neutron stars using a two-fluid description, where the outer and inner fluids give rise to two distinct fundamental frequencies. We confirm that the universal relation between the mass-scaled fundamental frequency of the outer-fluid-led mode and the stellar compactness, established previously for self-interacting dark matter admixed neutron stars, also holds in the mirror dark matter scenario. However, this universal relation becomes less robust when metric perturbations are included, compared with the corresponding results in the Cowling approximation. 
We further find that the inner-fluid-led fundamental frequency can also be expressed as a compactness-dependent relation that is largely independent of the normal matter equation of state, provided that the dark matter mass fraction is fixed. These results suggest that the simultaneous detection of the two fundamental frequencies could provide a way to constrain the dark matter mass fraction, even when the equation of state of normal matter remains uncertain. 
Finally, we find that the Cowling approximation estimates the fundamental frequency associated with the outer fluid with an accuracy comparable to that found for standard neutron stars without dark matter, while it performs even better for the frequency associated with the inner fluid.
\end{abstract}

\maketitle


\section{Introduction}
\label{sec:I}

Neutron stars, which are compact remnants formed through core-collapse supernova at the end of massive stellar evolution, provide a suitable environment for probing physics under extreme conditions far beyond those realized in the solar system~\cite{ST83}. For instance, constraining nuclear properties at high densities from terrestrial experiments is quite difficult because laboratory data mainly probe matter in the vicinity of nuclear saturation density, while mass measurements of massive neutron stars provide important constraints on the stiffness of the equation of state at higher densities. In fact, the discoveries of the massive neutron stars with masses around $2M_\odot$ have excluded soft equation of state whose predicted maximum masses do not reach the observed values~\cite{D10,A13,C20,F21,Romani22}. The observation of gravitational waves from the binary neutron star merger, GW170817~\cite{GW170817} constrained the dimensionless tidal deformability of neutron stars, which in turn led to an upper limit of $13.6$ km on the radius of a $1.40\ M_\odot$ neutron star~\cite{Annala18}. Furthermore, owing to the relativistic effect, photon trajectories are bent in the strong gravitational field around a neutron star. Since this bending is mainly governed by the strength of the gravitational field, which is characterized by the stellar compactness, \(M/R\), where \(M\) and \(R\) denote the stellar mass and radius, careful observations of pulsar light curves can provide direct constraints on the compactness, e.g.,~\cite{PFC83,LL95,PG03,PO14,SM18,Sotani20a}. Observationally, the masses and radii of PSR J0030+0451~\cite{Riley19,Miller19} and PSR J0740+6620~\cite{Riley21,Miller21} have been constrained by X-ray observations with the Neutron Star Interior Composition Explorer (NICER). These astronomical observations mainly constrain the equation of state in a relatively high-density region, whereas terrestrial experiments provide constraints in the lower-density regime~\cite{SNN22,SO22,SN23}. Combining constraints from astronomical observations and ground-based experiments is therefore essential for narrowing down the dense matter equation of state.

In addition to static properties, such as the mass, radius, and tidal deformability, the frequencies of stellar oscillations offer an independent way to probe neutron star properties. Since the oscillation frequencies strongly depend on the stellar properties, a relation between the frequencies and these stellar properties can be used inversely to extract stellar information from observed frequencies. This technique is known as {\textit{gravitational wave asteroseismology}}~\cite{AK1996,AK1998}, in analogy with terrestrial seismology and helioseismology. In fact, the quasi-periodic oscillations observed in the magnetar flares are often interpreted as torsional oscillations excited in neutron stars, and the neutron star mass, radius, and equation of state can be estimated by identifying the observed frequencies with specific torsional modes~\cite{GNHL2011,SNIO2012,SIO2016,SKS23,Sotani24a}. Gravitational waves from isolated neutron stars have not yet been observed, but their future detection could provide information on the stellar mass, radius, equation of state, and rotational properties, e.g.,~\cite{STM2001,SH2003,TL2005,SYMT2011,PA2012,DGKK2013,Sotani20b,Sotani21}. Furthermore, the technique of asteroseismology has also been applied to protoneutron stars formed immediately after core-collapse supernovae, both to investigate the explosion mechanism and/or to extract physical properties from supernova gravitational waves, e.g.,~\cite{FMP2003,FKAO2015,ST2016,SKTK2017,MRBV2018,TCPOF19,SS2019,ST2020,ST2020a,STT2021,SMT24,Sotani26}.

Since the oscillation frequencies from the neutron stars generally depend on model parameters, the extraction of stellar properties from these frequencies is also affected by the same model dependence. Namely, even if frequencies associated with stellar oscillations are detected through electromagnetic or gravitational-wave observations, it may still be difficult to extract meaningful information directly. To overcome this difficulty, universal relations are useful because they connect combinations of neutron star properties in a way that is insensitive to model parameters. So far, several universal (or empirical) relations combining oscillation frequencies with neutron star properties have been proposed, e.g.,~\cite{AK1998,TL2005,Chan14,SK21,HS26}.

Moreover, if dark matter is present inside or around neutron stars, it can also affect their macroscopic properties. Dark matter accounts for approximately $85\%$ of the total matter content in the universe~\cite{Aghanim20,Workman22}, and is expected to interact with normal matter only gravitationally or to have nongravitational interactions that are sufficiently weak. Owing to the strong gravitational field of neutron stars, dark matter may be captured and accumulate inside the star, leading to the formation of dark matter admixed neutron stars, although the capture mechanism and the relevant interaction cross sections remain uncertain. If such stars exist, their stellar properties depend on the underlying dark matter model, and observations may therefore provide a way to constrain the dark matter model inversely. In practice, a great deal of theoretical research on dark matter admixed neutron stars has been conducted so far by adopting different dark matter models, such as self-interacting fermionic
dark matter, e.g.,~\cite{Nelson19, Ivanytskyi20, Shawqi24, Rutherford25, KGS25, SK25}, mirror dark matter, e.g.,~\cite{Kain21, Berezhiani21, Emma22, KCSY26}, and dark matter interacting with nuclear matter through the Higgs channel, e.g.,~\cite{PL17, Das21, LLFD22, KS24, KS25}. We note that these models assume that dark matter does not self-annihilate. 
In fact, the presence of dark matter inside the neutron star may be probed via gravitational wave observations, e.g., Refs.~\cite{Ellis18a,Flores24}.

In the self-interacting fermionic dark matter model and the mirror dark matter model, dark matter is described as a separate fluid component from normal matter. That is, such systems should be discussed within a two-fluid approach. The two-fluid formalism for neutron stars was originally developed in the context of neutron star superfluidity, where uncharged matter, mainly the neutron fluid, is treated as a fluid distinct from charged matter, mainly the proton fluid~\cite{Comer99,Comer02,AC02,Comer03,Comer04}. When the oscillation frequencies of dark matter admixed neutron stars are considered within this two-fluid approach, additional frequencies associated with the dark matter component naturally appear, together with those associated with normal matter.

In the previous study, we found that, within the self-interacting fermionic dark matter model, the mass-scaled fundamental ($f$-) mode frequency associated with the outer fluid satisfies a universal relation with the stellar compactness, similar to the case of standard neutron stars without dark matter~\cite{SK25b}. This result was obtained in the Cowling approximation, where the metric is kept fixed during the fluid oscillations. Here, the outer fluid corresponds to normal matter in dark core configuration and to dark matter in dark halo configuration. On the other hand, even when the metric perturbations are included, this universality appears to hold to some extent in the mirror dark matter model, although a small deviation from the universal relation can be seen~\cite{KCSY26}. However, the equation of state dependence of this relation has not yet been discussed in calculations including metric perturbations. These results naturally raise the following question: does the small deviation from universality originate from the inclusion of metric perturbations, or from the adoption of the mirror dark matter model instead of the self-interacting dark matter model? In addition, the $f$-mode frequencies associated with the inner fluid may also exhibit a similar universal behavior. Furthermore, it remains unclear how accurately the Cowling approximation works for dark matter admixed neutron stars within the two-fluid approach. To clarify these issues, in this study, we examine the $f$-mode frequencies of dark matter admixed neutron stars in the mirror dark matter model, adopting various nuclear equations of state.

This manuscript is organized as follows. In Sec. \ref{sec:DMNS}, we briefly describe the equilibrium structures of dark matter admixed neutron stars adopting the mirror dark matter model, together with the nuclear equations of state used in this study. In Sec. \ref{sec:f-mode}, we examine the $f$-mode frequencies and discuss the relation between the mass-scaled $f$-mode frequencies and stellar compactness. Then, in Sec.~\ref{sec:Cowling}, we discuss the accuracy of the Cowling approximation for the $f$-mode frequencies within the two-fluid formalism. Finally, we summarize our findings in this study and the conclusions in Sec.~\ref{sec:Conclusion}. Unless otherwise mentioned, we adopt geometric units with $c=G=1$, where $c$ and $G$ denote the speed of light and the gravitational constant, respectively, and use the metric signature $(-,+,+,+)$.

\section{Equilibrium models of dark matter admixed neutron stars}
\label{sec:DMNS}

In this study, dark matter admixed neutron star models are constructed using a two-fluid approach. That is, the star contains both normal matter and dark matter, which are treated as two distinct fluid components, and the interaction between the two components is assumed to occur only through their common gravitational field. 
In particular, we first construct a static and spherically symmetric two-fluid stellar configuration, which serves as the unperturbed equilibrium model for the subsequent oscillation calculation. The spacetime is described by the metric
\begin{equation}
  ds^2 = -e^{2\Phi}dt^2 + e^{2\Lambda}dr^2 + r^2\left(d\theta^2 + \sin^2\theta d\phi^2\right), \label{eq:metric}
\end{equation}
where $\Phi$ and $\Lambda$ are metric functions that depend only on the radial coordinate, $r$. The metric function $\Lambda$ is related to the total gravitational mass function, $m(r)$ through $e^{-2\Lambda} = 1 - 2m/r$. Assuming that both normal matter and dark matter are described as perfect fluids, the energy-momentum tensor for each fluid component $x$ is given by
\begin{equation}
  T_{x}^{\mu\nu}=(\varepsilon_x+p_x)u_x^\mu u_x^\nu + p_xg^{\mu\nu}, \label{eq:Tx}
\end{equation}
where $\varepsilon_x$ and $p_x$ are the energy density and pressure of the fluid $x$, respectively, and $u_x^{\mu}$ is the four-velocity of the fluid $x$. The total energy-momentum tensor is obtained by summing the contribution from each fluid component, \(T_{\rm T}^{\mu\nu}=\sum_x T_x^{\mu\nu}\), and satisfies the conservation law \(\nabla_\mu T_{\rm T}^{\mu\nu}=0\). In the present model, we further assume that dark matter and normal matter do not exchange energy or momentum through any direct interaction other than gravity. Therefore, the energy-momentum tensor of each fluid component is separately conserved, i.e., \(\nabla_\mu T_x^{\mu\nu}=0\).

For the static stellar configuration, the four velocity of each fluid component is expressed as
\begin{equation}
   u_x^{\mu}=\left(e^{-\Phi},0,0,0\right).\label{eq:u_x} 
\end{equation}
Since the two fluids are at rest in the static background and share the same four-velocity, i.e., $u^\mu\equiv u_x^\mu$, the total energy-momentum tensor for the static stellar configuration can be written as
\begin{equation}
  T_{\rm T}^{\mu\nu}=(\varepsilon_{\rm T}+p_{\rm T})u^\mu u^\nu + p_{\rm T}g^{\mu\nu}, \label{eq:T_T}
\end{equation}
where $\varepsilon_{\rm T}$ and $p_{\rm T}$ are the total energy density and total pressure, respectively, given by 
\begin{equation}
  \varepsilon_{\rm T} = \sum_x\varepsilon_x\ \ \ {\rm and}\ \ \ p_{\rm T}=\sum_x p_x. \label{eq:total_ep}
\end{equation}
From the Einstein equations, together with the separate conservation law for each fluid component, one can obtain the Tolman-Oppenheimer-Volkoff (TOV) equations governing the equilibrium configuration in the two-fluid formalism as
\begin{align}
  m_x' &= 4\pi r^2\varepsilon_x, \label{eq:dm} \\
  p_x' &= -\frac{\left(4\pi r^3 p_{\rm T} +m\right)\left(\varepsilon_x+p_x\right)}{r(r-2m)}. \label{eq:dr} \\
  \Phi' &= \frac{4\pi r^3 p_{\rm T} +m}{r(r-2m)}, \label{eq:dPhi}
\end{align}
where $m_x$ denotes the mass function of the fluid component $x$, and the total gravitational mass function is given by $m(r)=\sum_x m_x(r)$~\cite{Goldman13,Sagun23}. Thus, while each fluid satisfies its own hydrostatic balance equation, the gravitational field is determined by the total energy density and total pressure of the two-fluid system. As usual, the equations of state for both normal matter and dark matter are required to integrate the TOV equations.

To examine the dependence on the nuclear equation of state, we adopt seven different equations of state. Two of them, DD2~\cite{DD2} and QMC-RMF4~\cite{QMC4}, are based on relativistic mean-field theory. Three equations of state, SLy4~\cite{SLy4}, SKa~\cite{SKa}, and SkI3~\cite{SkI3}, are constructed using Skyrme-type effective interactions. The Togashi equation of state is derived with the variational method~\cite{Togashi17}, while QHC21 is a quark-hadron crossover equation of state that adopts the Togashi equation of state for the hadronic matter sector~\cite{QHC21}. The nuclear saturation parameters for these equations of state and the maximum masses expected for neutron stars without dark matter are listed in Table~\ref{tab:EOS}. In this table, \(\eta\) denotes the parameter constructed from the incompressibility \(K_0\) and the slope parameter \(L\), defined as \(\eta=(K_0L^2)^{1/3}\), which is useful for characterizing the properties of low-mass neutron stars~\cite{SIOO14}. Constraints on the values of $K_0$ and $L$ are obtained from terrestrial experiments, e.g.,~\cite{Shlomo06,Garg18,Vinas14,Li19}. 
On the other hand, in this study, we focus on the mirror dark matter model. In this model, the dark sector is assumed to be an exact mirror copy of the Standard Model sector~\cite{Foot91}. Consequently, the dark matter equation of state can be taken to be identical to the normal matter equation of state.

\begin{table}
\caption{Nuclear saturation parameters \(K_0\), \(L\), and \(\eta=(K_0L^2)^{1/3}\) for the nuclear equations of state adopted in this study, together with the corresponding maximum mass \(M_{\rm max}/M_\odot\) of neutron stars without dark matter.
}
\label{tab:EOS}
\centering
\renewcommand{\arraystretch}{1.75}
\setlength{\tabcolsep}{8pt}
\begin{tabular}{ccccc}
\hline\hline
EOS & \(K_0\) (MeV) & \(L\) (MeV) & \(\eta\) \(MeV\) & \(M_{\rm max}/M_\odot\) \\
\hline
DD2 & 243 & 55.0 & 90.2 & 2.41 \\
QMC-RMF4 & 279 & 31.3 & 64.9 & 2.21 \\
SLy4 & 230 & 45.9 & 78.3 & 2.05 \\
SKa & 263 & 74.6 & 114 & 2.22 \\
SkI3 & 258 & 101 & 138 & 2.24 \\
Togashi & 245 & 38.7 & 71.6 & 2.21 \\
QHC21-BT & 245 & 38.7 & 71.6 & 2.20 \\
\hline\hline
\end{tabular}
\end{table}

To construct a stellar model within the two-fluid approach, one has to specify the central densities of both fluid components. Thus, when generating stellar sequences, for example, in the mass-radius relation, it is useful to fix an additional physical quantity, such as the ratio of the two central densities, e.g.,~\cite{KGS25} or the dark matter mass fraction, e.g.,~\cite{Rutherford25}. We note that, unlike standard neutron stars without dark matter, the stability of two-fluid configurations can not be discussed simply from the maximum mass obtained by varying the central density~\cite{KS25b}. In this study, we discuss stellar sequences with a fixed dark matter mass fraction, $M_{\rm DM}/M$, where $M$ denotes the total gravitational mass and $M_{\rm DM}$ denotes the dark matter contribution to the total mass. Depending on the radial distributions of the two fluids, the stellar models can be classified as dark core or dark halo configurations. Dark core and dark halo configurations are defined according to the relative sizes of the dark matter and normal matter regions. A dark core configuration corresponds to the case in which the radius of the dark matter region is smaller than that of the normal matter region, while a dark halo configuration corresponds to the opposite case. Since we adopt the mirror dark matter model in this study, the dark matter equation of state is identical to the normal matter equation of state. Therefore, stellar models with \(M_{\rm DM}/M<0.5\) correspond to dark core configurations, whereas those with \(M_{\rm DM}/M>0.5\) correspond to dark halo configurations. The case \(M_{\rm DM}/M=0.5\) is special: dark matter and normal matter contribute equally to the gravitational mass, and the radius of the dark matter region coincides with that of the normal matter region.

\begin{figure}[tbp]
\begin{center}
\includegraphics[scale=0.6]{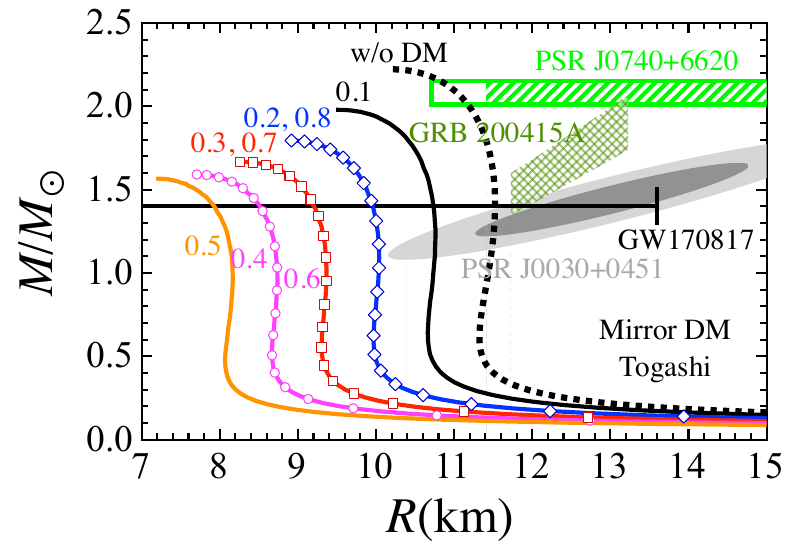} 
\end{center}
\caption{
Mass-radius relations of dark matter admixed neutron stars in the mirror dark matter model, constructed with the Togashi equation of state for normal matter. The total gravitational mass \(M/M_\odot\) is shown as a function of the stellar radius \(R\), defined as the outer radius of the two-fluid configuration, for fixed values of the dark matter mass fraction \(M_{\rm DM}/M\). The solid lines correspond to models with \(M_{\rm DM}/M=0.1\), 0.2, 0.3, 0.4, and 0.5, while the open circles, squares, and diamonds correspond to stellar models with \(M_{\rm DM}/M=0.6\), 0.7, and 0.8, respectively. For reference, the stellar sequence without dark matter is shown by the dotted line. Astronomical constraints from PSR J0030+0451~\cite{Riley19,Miller19}, PSR J0740+6620~\cite{Riley21,Miller21}, the \(1.40\,M_\odot\) neutron star radius inferred from GW170817~\cite{Annala18}, and GRB 200415A~\cite{SKS23} are also shown.
}
\label{fig:MR}
\end{figure}

As an example, Fig.~\ref{fig:MR} shows the mass-radius relations for stellar models constructed with the Togashi equation of state for normal matter, while varying the dark matter mass fraction $M_{\rm DM}/M$. Here, the stellar radius \(R\) is defined as the larger of the radii of the normal matter and dark matter. The solid lines correspond to stellar models with $M_{\rm DM}/M=0.1$, 0.2, 0.3, 0.4, and 0.5, while the open circles, squares, and diamonds correspond to models with $M_{\rm DM}/M=0.6$, 0.7, and 0.8, respectively. For reference, the stellar models without dark matter are shown with the dotted line, together with several constraints obtained from astronomical observations. 
The mirror dark matter setup has an exchange symmetry between the normal matter and dark matter components. Thus, for \(0<\chi<1\), a model with \(M_{\rm DM}/M=\chi\) and central densities \(\varepsilon_c^{\rm (DM)}=\varepsilon_1\), \(\varepsilon_c^{\rm (NM)}=\varepsilon_2\) has the same total mass and stellar radius as the model with \(M_{\rm DM}/M=1-\chi\) obtained by interchanging the two central densities, \(\varepsilon_c^{\rm (DM)}=\varepsilon_2\), \(\varepsilon_c^{\rm (NM)}=\varepsilon_1\). Here, \(\varepsilon_c^{\rm (DM)}\) and \(\varepsilon_c^{\rm (NM)}\) denote the central energy densities of dark matter and normal matter, respectively. This exchange symmetry explains why the mass-radius sequences for \(M_{\rm DM}/M=0.6\), 0.7, and 0.8 overlap with those for \(M_{\rm DM}/M=0.4\), 0.3, and 0.2, respectively, in Fig.~\ref{fig:MR}.

In this study, \(M_{\rm DM}/M\) is treated as a phenomenological parameter, and we consider a broad range of values to examine systematically how the oscillation spectrum depends on the dark matter content. The equilibrium and perturbation formalisms adopted here do not specify how the dark matter component is accumulated or formed and neglect nongravitational interactions between the two fluids. The astrophysically realized value of \(M_{\rm DM}/M\) therefore depends on the dark matter candidate, its possible interactions with normal matter and within the dark sector, the ambient dark matter density and velocity distribution, the stellar age, and the formation history~\cite{Grippa}.

In the standard capture scenario, halo dark matter becomes gravitationally bound to a neutron star after losing sufficient kinetic energy through scattering with the stellar constituents. For self-annihilating WIMPs (weakly interacting massive particles), the retained population can be further limited by the establishment of capture–annihilation equilibrium. Asymmetric or otherwise nonannihilating dark matter avoids this depletion and can therefore accumulate more efficiently for the same capture rate, while dark matter self-interactions or a larger ambient density may further enhance the accumulation. Nevertheless, for representative local-halo densities, \(\rho_{\rm DM}\simeq 0.3-0.5\) GeV cm\(^{-3}\), the dark matter mass accumulated over a neutron star lifetime is generally far too small to modify its bulk structure or oscillation spectrum~\cite{Salas21,Grippa,Ellis18,Jungman96}. Denser environments, such as the Galactic center or compact dark matter overdensities, can increase the capture rate, but conventional scattering capture is still not generally expected to produce \(M_{\rm DM}/M \sim 0.1\).

Macroscopic dark matter fractions therefore require nonstandard formation or production channels. Possible examples include the co-formation of the normal and dark components from an initial overdensity, the formation of a compact object in a dissipative dark sector followed by baryonic accretion, the dynamical settling of a dense dark matter cloud around a pre-existing neutron star, or the production or conversion of dark degrees of freedom within the star~\cite{Ellis18, Capela13}. These mechanisms are highly model dependent, and current mass–radius and gravitational-wave observations constrain \(M_{\rm DM}/M\) only within a specified model for the normal matter and dark matter equations of state and their distributions. The broad range considered here should therefore be regarded as a systematic exploration of the equilibrium and oscillation properties permitted by the two-fluid model, rather than as a prediction of standard halo capture. Moreover, owing to the exchange symmetry discussed above, configurations with \(M_{\rm DM}/M > 1/2\) are more naturally interpreted as mirror matter-dominated stars containing a subdominant normal matter component.

\section{Relation between mass-scaled \(f\)-mode frequencies and stellar compactness}
\label{sec:f-mode}

We now consider stellar oscillations on the equilibrium models described in the previous section. To systematically examine the dependence on the model parameters, we first adopt the Cowling approximation for determining the oscillation frequencies, in which the metric perturbations are neglected. Although the determination of oscillation frequencies including metric perturbations in the two-fluid approach has recently become possible~\cite{KCSY26}, the required computational time/cost remains very high, making systematic parameter survey a daunting task. With the Cowling approximation, the perturbation equations in the two-fluid approach can be derived from the linearized energy-momentum conservation law for each fluid component ($x$), i.e., $\nabla_\mu\left(\delta T_x^{\mu\nu}\right)=0$~\cite{SK25}. 
The oscillation problem is then solved by imposing regularity at the stellar center and requiring the Lagrangian pressure perturbation of each fluid component to vanish at its own surface. The oscillation problem eventually reduces to an eigenvalue problem for the angular frequency \(\omega\), from which the ordinary frequency is obtained as \(f=\omega/(2\pi)\). The explicit form of the perturbation equations and boundary conditions is given in Ref.~\cite{SK25}. In this study, we focus only on the $\ell=2$ oscillation modes, because they are expected to be the dominant contributors to gravitational-wave radiation.

\begin{figure}[tbp]
\begin{center}
\includegraphics[scale=0.6]{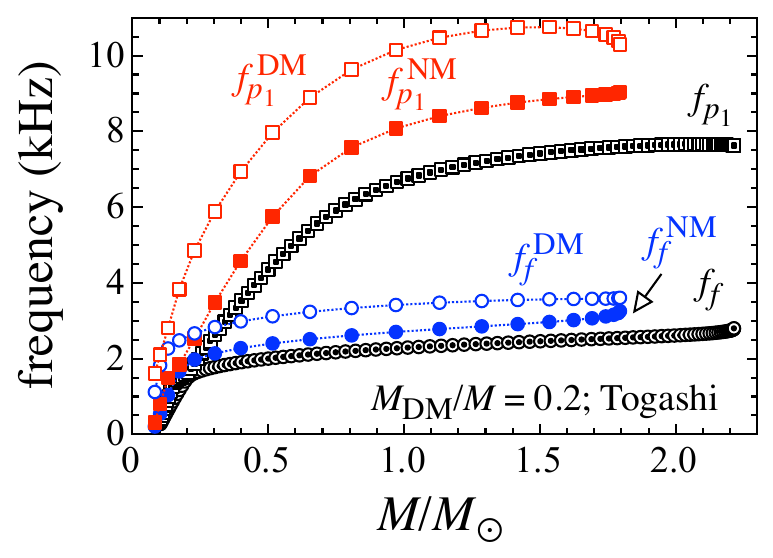} 
\end{center}
\caption{
The $f$- and $p_1$-mode frequencies associated with the normal matter component (filled symbols) and the dark matter component (open symbols) are shown as functions of the total gravitational mass. The stellar models are constructed in the mirror dark matter model with \(M_{\rm DM}/M=0.2\), using the Togashi equation of state for both fluid components. For reference, the corresponding frequencies of the neutron star models without dark matter are also shown by double symbols. 
}
\label{fig:MDF02_Togashi}
\end{figure}

To examine the behavior of the frequencies, Fig.~\ref{fig:MDF02_Togashi} shows the fundamental ($f$-) and the first pressure ($p_1$-) mode frequencies associated with normal matter and dark matter, denoted by $f_f^{x}$ and $f_{p_1}^{x}$, respectively, where $x=\mathrm{NM}$ for normal matter and $x=\mathrm{DM}$ for dark matter. These frequencies are shown as functions of the total gravitational mass \(M\) for stellar models with \(M_{\rm DM}/M=0.2\), constructed using the Togashi equation of state for both fluid components. For reference, the $f$- and $p_1$-mode frequencies of neutron star models without dark matter are also shown by the double circles and double squares, respectively. Since $M_{\rm DM}/M=0.2<0.5$, all stellar models considered here correspond to dark core configurations. As shown in Fig.~\ref{fig:MDF02_Togashi}, for stars of the same total mass, the frequencies associated with the inner fluid are higher than those associated with the outer fluid. In addition, even the $f$-mode frequencies associated with normal matter differ from the $f$-mode frequencies of the neutron star models without dark matter, owing to the presence of the dark matter component.

\begin{figure}[tbp]
\begin{center}
\includegraphics[scale=0.6]{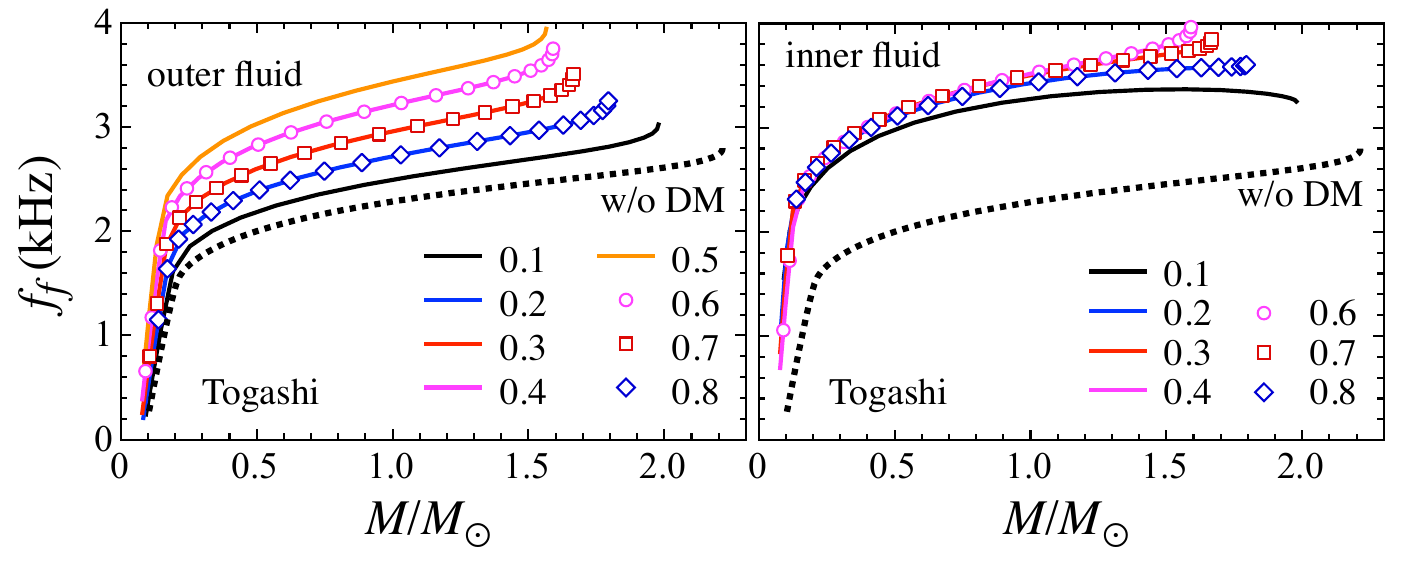} 
\end{center}
\caption{
The $f$-mode frequencies, $f_f$, are shown as functions of the total gravitational mass \(M\) for stellar models constructed with the Togashi equation of state for different dark matter mass fractions \(M_{\rm DM}/M\), adopting the mirror dark matter model. The left and right panels show the \(f\)-mode frequencies associated with the outer and inner fluids, respectively. The solid lines correspond to models with \(M_{\rm DM}/M=0.1\), 0.2, 0.3, 0.4, and 0.5, while the open circles, squares, and diamonds correspond to models with \(M_{\rm DM}/M=0.6\), 0.7, and 0.8, respectively. For reference, the \(f\)-mode frequencies of neutron star models without dark matter are also shown by the dotted line.
}
\label{fig:ff_Togashi}
\end{figure}

In the left and right panels of Fig.~\ref{fig:ff_Togashi}, the $f$-mode frequencies associated with the outer and inner fluid are shown as functions of the total gravitational mass for stellar models constructed with the Togashi equation of state, while varying the dark matter mass fraction. For reference, the $f$-mode frequencies of stellar models without dark matter constructed with the same Togashi equation of state are also shown by the dotted line. As mentioned in the connection with Fig.~\ref{fig:MR}, for example, the dark core configurations with $M_{\rm DM}/M=0.2$ have the same equilibrium structure as the dark halo configurations with $M_{\rm DM}/M=0.8$ after exchanging the dark matter and normal matter components. In these two cases, the outer fluid corresponds to normal matter for \(M_{\rm DM}/M=0.2\), while it corresponds to dark matter for \(M_{\rm DM}/M=0.8\). Thus, as expected from the exchange symmetry, the $f$-mode frequencies associated with the outer fluid in the stellar models with $M_{\rm DM}/M=\chi$, for $0<\chi<1$, are identical to those in the corresponding models with $M_{\rm DM}/M=1-\chi$. The same correspondence also holds for the $f$-mode frequencies associated with the inner fluid, as seen in Fig.~\ref{fig:ff_Togashi}. 
In addition, we find that the $f$-mode frequencies associated with the outer fluid strongly depend on $M_{\rm DM}/M$, while the dependence of those with the inner fluid on $M_{\rm DM}/M$ seems to be weak. 
However, as shown below, this trend is reversed when the mass-scaled \(f\)-mode frequencies are considered.

\begin{figure}[tbp]
\begin{center}
\includegraphics[scale=0.6]{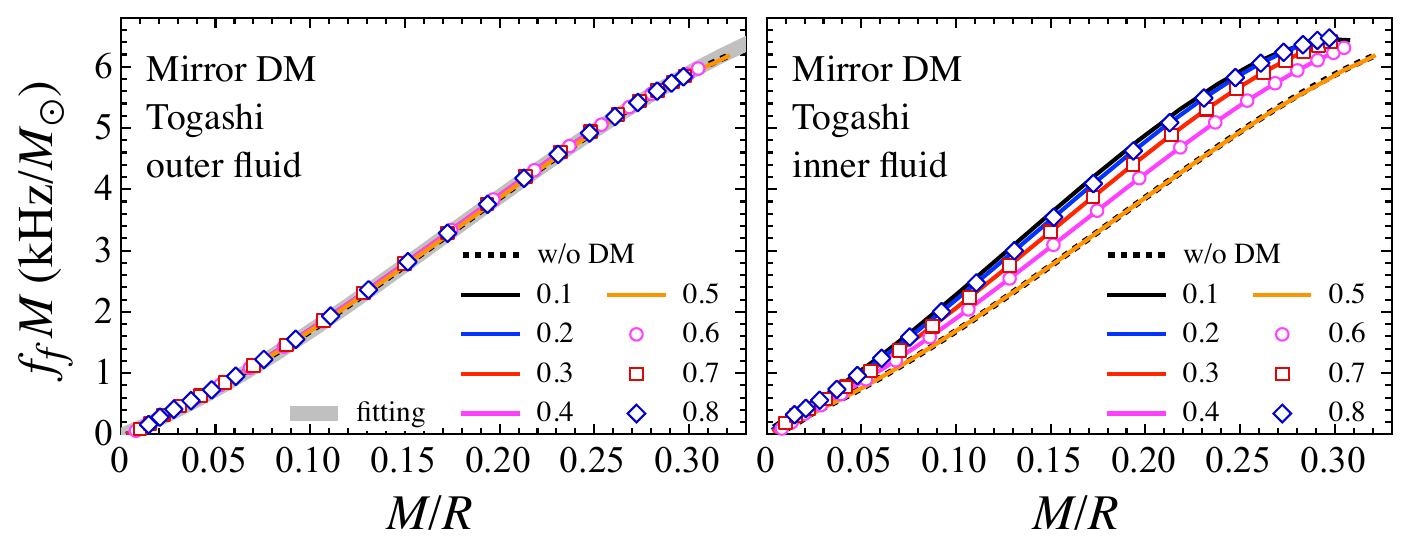} 
\end{center}
\caption{
The mass-scaled $f$-mode frequencies are shown as functions of the stellar compactness for dark matter admixed neutron stars with fixed dark matter mass fractions, adopting the mirror dark matter model. The stellar models are constructed with the Togashi equation of state. The left and right panels correspond to the frequencies associated with the outer and inner fluids, respectively. The dotted line denotes the results for neutron star models without dark matter constructed with the Togashi equation of state, while the thick solid line denotes the universal relation for neutron stars without dark matter.
}
\label{fig:ffoi_Togashi}
\end{figure}

For neutron stars without dark matter, the \(f\)-mode frequencies generally depend on the equation of state, but the relation between the mass-scaled \(f\)-mode frequency and the stellar compactness is known to be almost independent of the equation of state~\cite{TL2005}. In practice, using the $f$-mode frequencies computed with the Cowling approximation, the relation is expressed as
\begin{equation}
  f_f^{\rm (out)} M\ {\rm (kHz}/M_\odot) = -0.01932 + 2.115\tilde{\cal C} + 1.731\tilde{\cal C}^2 - 0.5720\tilde{\cal C}^3, \label{eq:ff_out}
\end{equation}
where $\tilde{\cal C}$ is the stellar compactness normalized by 0.172, i.e., $\tilde{\cal C}\equiv (M/R)/0.172$~\cite{SK25}. Here, 0.172 corresponds to $M/R$ for a canonical neutron star with mass ($1.40\, M_\odot$) and radius (12 km). For dark matter admixed neutron stars, the \(f\)-mode frequencies depend not only on the nuclear equation of state, but also on the dark matter model, the dark matter mass fraction, and the parameters characterizing the dark matter sector. Nevertheless, in our previous study based on the self-interacting dark matter model, we found that the universal relation given by Eq. (\ref{eq:ff_out}) still holds even for the mass-scaled \(f\)-mode frequency associated with the outer fluid independently of the nuclear equation of state, the dark matter mass fraction, and the dark matter model parameters~\cite{SK25b}. On the other hand, as shown in the left panel of Fig.~\ref{fig:ffoi_Togashi}, we confirm that the universal relation for the outer fluid $f$-mode also holds in the mirror dark matter model, even though this model is different from the self-interacting dark matter model. This indicates that the slight weakening of the universality found in calculations including metric perturbations~\cite{KCSY26} originates from the inclusion of metric perturbations, rather than from the change of the dark matter model. Meanwhile, as seen in the right panel of Fig.~\ref{fig:ffoi_Togashi}, the mass-scaled $f$-mode frequencies associated with the inner fluid depend on $M_{\rm DM}/M$. 
In particular, the dependence on \(M_{\rm DM}/M\) becomes increasingly pronounced as the dark matter mass fraction \(M_{\rm DM}/M\) increases up to 0.5.

\begin{figure}[tbp]
\begin{center}
\includegraphics[scale=0.5]{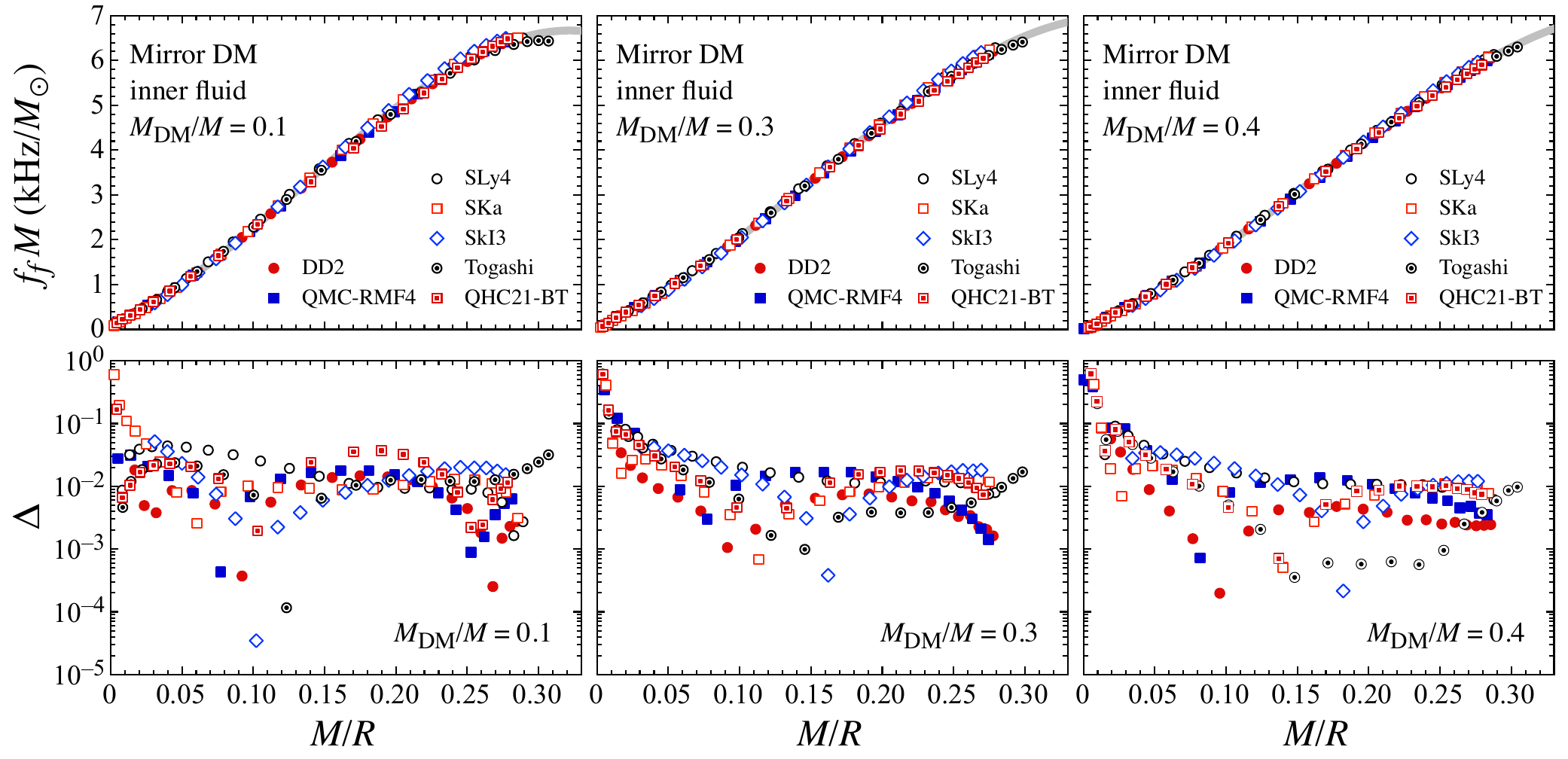} 
\end{center}
\caption{
Top panels: mass-scaled $f$-mode frequencies associated with the inner fluid are shown as a function of total stellar compactness $M/R$, adopting various nuclear equations of state. The left, middle, and right panels correspond to stellar models with \(M_{\rm DM}/M=0.1\), 0.3, and 0.4, respectively. The thick solid line in each panel denotes the fitting relation given by Eq.~(\ref{eq:ff_in}), with coefficients listed in Table~\ref{tab:ffi_coeffi}. Bottom panels: relative deviations of the mass-scaled \(f\)-mode frequencies obtained with different nuclear equations of state from the corresponding fitting relation.
}
\label{fig:MDMi_comp}
\end{figure}

\begin{table}
\caption{Fitting coefficients \(a_i\) with \(i=0-3\) in Eq.~(\ref{eq:ff_in}) for different values of the dark matter mass fraction \(M_{\rm DM}/M\).
} 
\label{tab:ffi_coeffi}
\centering
\renewcommand{\arraystretch}{1.75}
\setlength{\tabcolsep}{9pt}
\begin{tabular}{ccccc}
\hline\hline
\(M_{\rm DM}/M\) & \(a_0\) & \(a_1\) & \(a_2\) & \(a_3\) \\
\hline
0.1 & 0.1075 & 2.350 & 3.104 & \(-1.3282\) \\
0.2 & 0.07959 & 2.392 & 2.765 & \(-1.1332\) \\
0.3 & 0.05114 & 2.344 & 2.420 & \(-0.9345\) \\
0.4 & 0.01746 & 2.268 & 2.036 & \(-0.7294\) \\
0.5 & \(-0.01932\) & 2.115 & 1.731 & \(-0.5720\) \\
\hline\hline
\end{tabular}
\end{table}

Although the relation between the mass-scaled $f$-mode frequencies associated with the inner fluid and stellar compactness depends on the dark matter mass fraction, we find that, for a fixed value of $M_{\rm DM}/M$, this relation is almost independent of the equation of state for nuclear matter. As shown in Fig.~\ref{fig:MDMi_comp}, for a fixed value of \(M_{\rm DM}/M\), the mass-scaled \(f\)-mode frequency associated with the inner fluid is almost independent of the choice of nuclear equation of state. This relation can be expressed as
\begin{equation}
  f_f^{\rm (in)} M\ {\rm (kHz}/M_\odot) = a_0 + a_1\tilde{\cal C} + a_2\tilde{\cal C}^2 + a_3\tilde{\cal C}^3, \label{eq:ff_in}
\end{equation}
where $a_i$ with $i=0-3$ are the fitting coefficients that depend on $M_{\rm DM}/M$. The coefficients in Eq.~(\ref{eq:ff_in}) are listed in Table~\ref{tab:ffi_coeffi}. We note that the coefficients in Eq.~(\ref{eq:ff_in}) for the stellar model with $M_{\rm DM}/M=0.5$ should be equal to those in Eq. (\ref{eq:ff_out}).
In addition, as mentioned above, the stellar models with $M_{\rm DM}/M = \chi$ are the same as those with $M_{\rm DM}/M = 1 - \chi$ due to the nature of the mirror dark matter model, leading to the fact that the frequencies excited in the stellar models with $M_{\rm DM}/M (= \chi) > 0.5$ become completely the same as those with $M_{\rm DM}/M = 1 - \chi$. This is the reason why we only show the results for $M_{\rm DM}/M \le 0.5$ in Table~\ref{tab:ffi_coeffi}. 
This result suggests that, if two distinct \(f\)-mode frequencies associated with the outer and inner fluids are observed, their comparison with the corresponding universal relations may provide a way to infer the dark matter mass fraction in dark matter admixed neutron stars. Additionally, Fig.~\ref{fig:MDMi_comp} shows that the equation-of-state-insensitive relation becomes slightly tighter as \(M_{\rm DM}/M\) increases. This suggests that the inner-fluid \(f\)-mode frequency may provide a robust indicator of the dark matter mass fraction, at least within the Cowling approximation. A more complete assessment, however, requires calculations including metric perturbations, which will be necessary to verify whether this fixed-\(M_{\rm DM}/M\) universality persists beyond the Cowling approximation.

\section{Accuracy of the Cowling approximation for \(f\)-mode frequencies}
\label{sec:Cowling}

\begin{figure}[tbp]
\begin{center}
\includegraphics[scale=0.6]{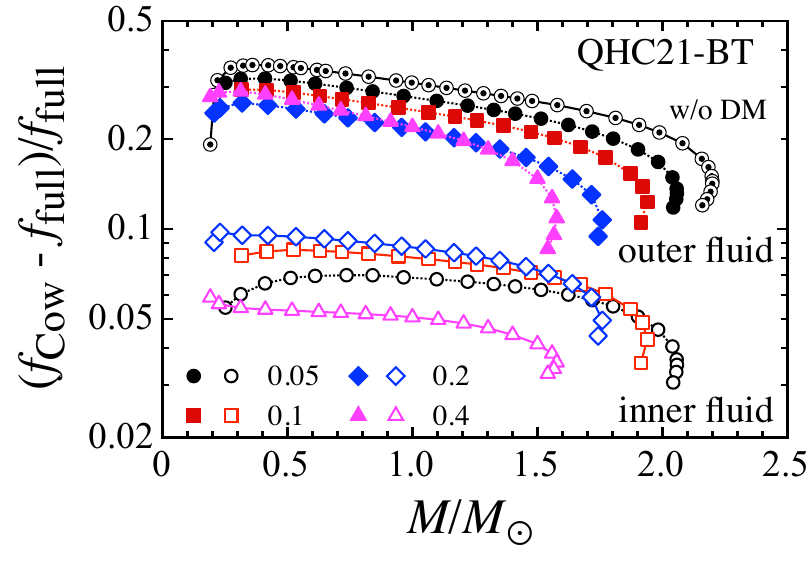} 
\end{center}
\caption{
Relative deviation of the $f$-mode frequencies determined with the Cowling approximation, $f_{\rm Cow}$, from those with full perturbations, $f_{\rm full}$, is shown as a function of the total gravitational mass of the neutron stars with various dark matter mass fractions, $M_{\rm DM}/M=0.05$, 0.1, 0.2, and 0.4, adopting the QHC21-BT EOS for normal matter. The filled and open marks denote the $f$-mode frequencies associated with the outer and inner fluid, while the double-circles denote the result for neutron stars without dark matter. 
}
\label{fig:Delta_ff}
\end{figure}

We now discuss the accuracy of the Cowling approximation for dark matter admixed neutron stars within the two-fluid formalism. For standard neutron stars without dark matter, the Cowling approximation can reproduce the qualitative behavior of fluid oscillation frequencies, even though metric perturbations are neglected. 
Although the Cowling approximation can lead to relative deviations of up to approximately \(30\%\), its accuracy is known to improve with increasing \(\ell\) and for higher-order modes~\cite{YK97,ST2020}. 
To assess its accuracy for dark matter admixed neutron stars in the two-fluid approach, we compare the frequencies obtained in the Cowling approximation with those computed including metric perturbations in Ref.~\cite{KCSY26}. We note that Ref.~\cite{KCSY26} neglects the damping rate of the complex eigenfrequency, i.e., it uses only the real part of the quasinormal-mode frequency. This assumption is well justified for the \(f\)- and \(p_1\)-modes of neutron stars without dark matter~\cite{Sotani22}, because the damping rates of fluid-dominated oscillation modes are generally much smaller than their oscillation frequencies. In Fig.~\ref{fig:Delta_ff}, we show the relative deviation of the $f$-mode frequencies obtained with the Cowling approximation, $f_{\rm Cow}$, from those with metric perturbations, $f_{\rm full}$, for stellar models constructed with QHC21-BT equation of state, adopting the mirror dark matter model. The filled and open symbols correspond to the \(f\)-mode frequencies associated with the outer and inner fluids, respectively. For reference, the results for the stellar models without dark matter are also shown by double circles. From this figure, one can see that the relative deviation for the \(f\)-mode frequencies associated with the outer fluid is comparable to that for stellar models without dark matter, although the deviation becomes slightly smaller as dark matter mass fraction \(M_{\rm DM}/M\) increases. On the other hand, the relative deviation for the \(f\)-mode frequencies associated with the inner fluid is much smaller than both that for the outer-fluid frequencies and that for neutron star models without dark matter. This indicates that the Cowling approximation reproduces the inner-fluid mode frequencies with significantly better accuracy. We note that the frequencies are always an overestimate with the Cowling approximation.
In any case, although we compared the frequencies only with the QHC21-BT equation of state, it is better to additionally assess the accuracy of the Cowling approximation with several other equations of state in the future.

\section{Conclusion}
\label{sec:Conclusion}

Dark matter may be accumulated inside neutron stars through capture or gravitational confinement, leading to the possible formation of dark matter admixed neutron stars. If such objects exist, their macroscopic properties and oscillation spectra can differ from those of standard neutron stars composed only of normal matter. Gravitational waves from dark matter admixed neutron stars therefore provide a possible observational channel for distinguishing these objects and extracting information about the dark matter component. In particular, if dark matter interacts with normal matter only through gravity, a dark matter admixed neutron star should be described within a two-fluid approach. Such a system naturally supports two distinct \(f\)-mode oscillations, associated with the two fluid components. In this study, we confirmed that the universal relation between the mass-scaled $f$-mode frequency associated with the outer fluid and the stellar compactness also holds in the mirror dark matter model, independently of the normal matter equation of state and the dark matter mass fraction. At the same time, comparison with calculations including metric perturbations shows that this universality becomes less robust beyond the Cowling approximation. On the other hand, we also showed that, for a fixed dark matter mass fraction, the mass-scaled $f$-mode frequency associated with the inner fluid can be well expressed as a function of the stellar compactness, independently of the normal matter equation of state. 
That is, careful observations of two distinct \(f\)-mode frequencies may allow one to constrain the dark matter mass fraction of a dark matter admixed neutron star, even when the normal matter EOS remains uncertain. 
Finally, we estimated the accuracy of the Cowling approximation for the $f$-mode frequencies of dark matter admixed neutron stars within the two-fluid approach. For the \(f\)-mode frequencies associated with the outer fluid, the Cowling approximation gives relative deviations comparable to those found for neutron stars without dark matter, namely within \(\sim 30\%\). In contrast, the frequencies associated with the inner fluid are reproduced more accurately, with relative deviations within \(\sim 10\%\). In both cases, the accuracy of the Cowling approximation improves as the stellar mass increases. These results provide a useful benchmark for future fully relativistic studies of two-fluid oscillations and for assessing the reliability of dark matter signatures in neutron star asteroseismology.

In this study, focusing on the frequencies excited in the dark matter admixed neutron stars, the oscillation frequencies were determined through linear perturbation analysis. Thus, from this study, we cannot discuss the amplitude of each oscillation mode, which leads to the radiation energy for each excited oscillation mode. In addition, in this study, we showed that the frequencies excited in neutron stars with canonical mass become more than a few kHz, similar to those of neutron stars without dark matter. So, unfortunately, these modes are almost impossible to detect with current gravitational wave detectors, while the next-generation detectors, such as the Einstein Telescope or Cosmic Explorer, may be able to detect them if the gravitational wave energy is sufficiently large.

\acknowledgments

This work is supported in part by the Mitsubishi Foundation through grant No. 202510029.





\begin{thebibliography}{999}

\bibitem{ST83}
   S. L. Shapiro and S. A. Teukolsky, {\it Black Holes, White Dwarfs, and Neutron Stars: The Physics of Compact Objects}  (Wiley-Interscience, New York, 1983).

\bibitem{D10} 
   P. Demorest, T. Pennucci, S. Ransom, M. Roberts, and J. Hessels, Nature {\bf 467}, 1081 (2010).

\bibitem{A13} 
   J. Antoniadis {\it et al.}, Science {\bf 340}, 6131 (2013).

\bibitem{C20}    
   H. T. Cromartie {\it et al.}, Nature Astronomy {\bf 4}, 72 (2020).

\bibitem{F21}    
   E. Fonseca {\it et al.}, Astrophys. J. {\bf 915}, L12 (2021).

\bibitem{Romani22} 
   R. W. Romani, D. Kandel, A. V. Filippenko, T. G. Brink, and W. Zheng, Astrophys. J. {\bf 934}, L17 (2022).

\bibitem{GW170817}  
   B. P. Abbott {\it et al}. (LIGO Scientific and Virgo Collaborations), Phys. Rev. Lett. {\bf 119}, 161101 (2017).

\bibitem{Annala18}  
   E. Annala, T. Gorda, A. Kurkela, and A. Vuorinen, Phys. Rev. Lett. {\bf 120}, 172703 (2018).

 
\bibitem{PFC83} 
   K. R. Pechenick, C. Ftaclas, and J. M. Cohen, Astrophys. J. {\bf 274}, 846 (1983).

\bibitem{LL95} 
   D. A. Leahy and L. Li, Mon. Not. R. Astron. Soc. {\bf 277}, 1177 (1995).

\bibitem{PG03} 
   J. Poutanen and M. Gierlinski, Mon. Not. R. Astron. Soc. {\bf 343}, 1301 (2003).

\bibitem{PO14} 
   D. Psaltis and F. \"{O}zel, Astrophys. J. {\bf 792}, 87 (2014). 

\bibitem{SM18} 
   H. Sotani and U. Miyamoto, Phys. Rev. D {\bf 98}, 044017 (2018); {\bf 98}, 103019 (2018).

\bibitem{Sotani20a} 
   H. Sotani, Phys. Rev. D {\bf 101}, 063013 (2020).

\bibitem{Riley19} 
   T. E. Riley {\it et al.}, Astrophys. J.  {\bf 887}, L21 (2019).
  
\bibitem{Miller19} 
   M. C. Miller {\it et al.}, Astrophys. J.  {\bf 887}, L24 (2019).
   
\bibitem{Riley21} 
   T. E. Riley {\it et al.}, Astrophys. J.  {\bf 918}, L27 (2021).
   
\bibitem{Miller21} 
   M. C. Miller {\it et al.}, Astrophys. J.  {\bf 918}, L28 (2021).


\bibitem{SNN22}
   H. Sotani, N. Nishimura, and T. Naito, Prog. Theor. Exp. Phys. {\bf 2022}, 041D01 (2022).

\bibitem{SO22}
   H. Sotani and S. Ota, Phys. Rev. D {\bf 106}, 103005 (2022).

\bibitem{SN23}
   H. Sotani and T. Naito, Phys. Rev. C {\bf 107}, 035802 (2023).

\bibitem{AK1996}
   N. Andersson and K. D. Kokkotas, Phys.\ Rev.\ Lett.\ {\bf 77}, 4134 (1996).

\bibitem{AK1998}
   N. Andersson and K. D. Kokkotas, Mon.\ Not.\ R. Astron.\ Soc.\ {\bf 299}, 1059 (1998).



\bibitem{GNHL2011}
   M. Gearheart, W. G. Newton, J. Hooker, and B. -A. Li, Mon. Not. R. Astron. Soc. {\bf 418}, 2343 (2011).
  
\bibitem{SNIO2012}
   H. Sotani, K. Nakazato, K. Iida, and K. Oyamatsu, Phys. Rev. Lett. {\bf 108}, 201101 (2012);
   Mon. Not. R. Astron. Soc. {\bf 428}, L21 (2013); {\bf 434}, 2060 (2013).

\bibitem{SIO2016}
   H. Sotani, K. Iida, and K. Oyamatsu, New Astron. {\bf 43}, 80 (2016);
   Mon. Not. R. Astron. Soc. {\bf 464}, 3101 (2017); {\bf 479}, 4735 (2018); 
   {\bf 489}, 3022 (2019).

\bibitem{SKS23}
   H. Sotani, K. D. Kokkotas, and N. Stergioulas, Astron. Astrophys. {\bf 676}, A65 (2023).

\bibitem{Sotani24a}  
   H. Sotani, Universe {\bf 10}, 231 (2024)


 
\bibitem{STM2001}
   H. Sotani, K. Tominaga, and K. I. Maeda, Phys.\ Rev.\ D {\bf 65}, 024010 (2001).

\bibitem{SH2003}
   H. Sotani and T. Harada, Phys.\ Rev.\ D {\bf 68}, 024019 (2003);
   H. Sotani, K. Kohri, and T. Harada, {\it ibid}.\ {\bf 69}, 084008 (2004).

\bibitem{TL2005}
   L. K. Tsui and P. T. Leung, Mon.\ Not.\ R. Astron.\ Soc.\ {\bf 357}, 1029 (2005).

\bibitem{SYMT2011}
   H. Sotani, N. Yasutake, T. Maruyama, and T. Tatsumi, Phys.\ Rev.\ D {\bf 83} 024014 (2011).

\bibitem{PA2012}
   A. Passamonti and N. Andersson, Mon.\ Not.\ R. Astron.\ Soc.\ {\bf 419}, 638 (2012).

\bibitem{DGKK2013}
   D. D. Doneva, E. Gaertig, K. D. Kokkotas, and C. Kr\"{u}ger, Phys.\ Rev.\ D {\bf 88}, 044052 (2013).

\bibitem{Sotani20b}
   H. Sotani, Phys. Rev. D {\bf 102}, 063023 (2020); 103021 (2020).

\bibitem{Sotani21}
   H. Sotani, Phys. Rev. D {\bf 103}, 123015 (2021).

   


\bibitem{FMP2003}
   V. Ferrari, G. Miniutti, and J. A. Pons, Mon. Not. R. Astron. Soc. {\bf 342}, 629 (2003).

\bibitem{FKAO2015}
   J. Fuller, H. Klion, E. Abdikamalov, and C. D. Ott, Mon.\ Not.\ R. Astron.\ Soc.\ {\bf 450}, 414 (2015).

\bibitem{ST2016}
   H. Sotani and T. Takiwaki, Phys.\ Rev.\ D {\bf 94}, 044043 (2016); {\bf 102}, 023028 (2020).
   
\bibitem{SKTK2017}
   H. Sotani, T. Kuroda, T. Takiwaki, and K. Kotake, Phys.\ Rev.\ D {\bf 96}, 063005 (2017); {\bf 99}, 123024 (2019).

\bibitem{MRBV2018}
  V. Morozova, D. Radice, A. Burrows, and D. Vartanyan, Astrophys. J. {\bf 861}, 10 (2018).


\bibitem{TCPOF19}
   A. Torres-Forn\'{e}, P. Cerd\'{a}-Dur\'{a}n, A. Passamonti, M. Obergaulinger, and J. A. Font, Mon. Not. R. Astron. Soc. {\bf 482}, 3967 (2019).

\bibitem{SS2019}
   H. Sotani and K. Sumiyoshi, Phys.\ Rev.\ D {\bf 100}, 083008 (2019); Mon. Not. R. Astron. Soc. {\bf 507}, 2766 (2021).

\bibitem{ST2020} 
   H. Sotani and T. Takiwaki, Phys.\ Rev.\ D {\bf 102}, 063025 (2020).

\bibitem{ST2020a}
   H. Sotani and T. Takiwaki, Mon. Not. R. Astron. Soc. {\bf 498}, 3503 (2020).

\bibitem{STT2021}
   H. Sotani, T. Takiwaki, and H. Togashi, Phys.\ Rev.\ D {\bf 104}, 123009 (2021).

\bibitem{SMT24}
   H. Sotani, B. M\"{u}ller, and T. Takiwaki, Phys.\ Rev.\ D {\bf 109}, 123021 (2024); {\bf 112}, 083018 (2025).

\bibitem{Sotani26}
   H. Sotani, Class. Quantum Gravity {\bf 43}, 093001 (2026).

\bibitem{Chan14}
   T. K. Chan, Y.-H. Sham, P. T. Leung, L.-M. Lin, Phys. Rev. D {\bf 90}, 124023 (2014).

\bibitem{SK21}
   H. Sotani and B. Kumar, Phys. Rev. D {\bf 104}, 123002 (2021).

\bibitem{HS26}
   H. Sotani, Phys. Rev. D {\bf 113}, 023017 (2026).

\bibitem{Aghanim20}
   N. Aghanim et al. (Planck Collaboration), Astron. Astrophys. {\bf 641}, A6 (2020).
   
\bibitem{Workman22}
   R. L. Workman et al. (Particle Data Group), Prog. Theor. Exp. Phys. {\bf 2022}, 083C01 (2022).

\bibitem{Nelson19}
   A. E. Nelson, S. Reddy, and D. Zhou, J. Cosmol. Astropart. Phys. {\bf 07} 012 (2019).

\bibitem{Ivanytskyi20}
   O. Ivanytskyi, V. Sagun, and I. Lopes, Phys. Rev. D {\bf 102}, 063028 (2020).
   
\bibitem{Shawqi24}
   S. Shawqi and S. M. Morsink, Astrophys. J. 975, 123 (2024).
   
\bibitem{Rutherford25}
   N. Rutherford, C. Prescod-Weinstein, and A. Watts, Phys. Rev. D {\bf 111}, 123034 (2025).
   
\bibitem{KGS25}
   A. Kumar, S. Girmohanta, and H. Sotani, Eur. Phys. J. C {\bf 85}, 1109, (2025).

\bibitem{SK25}
   H. Sotani and A. Kumar, Phys. Rev. D {\bf} 111, 123013 (2025).

\bibitem{Kain21}
   B. Kain, Phys. Rev. D {\bf 103}, 043009 (2021).

\bibitem{Berezhiani21}
   Z. Berezhiani, R. Biondi, M. Mannarelli, and F. Tonelli, Eur. Phys. J. C {\bf 81}, 1036 (2021).

\bibitem{Emma22}
   M. Emma, F. Schianchi, F. Pannarale, V. Sagun, T. Dietrich, Particles {\bf 5}, 273 (2022).

\bibitem{KCSY26}
   A. Kumar, D. A. Caballero, H. Sotani, N. Yunes, arXiv:2605.03305 [gr-qc].

\bibitem{PL17}
   G. Panotopoulos and I. Lopes, Phys. Rev. D {\bf 96}, 083004 (2017).

\bibitem{Das21}
   H. C. Das, A. Kumar, and S. K. Patra, Phys. Rev. D {\bf 104}, 063028 (2021).

\bibitem{LLFD22}
   O. Lourenço, C. H. Lenzi, T. Frederico, and M. Dutra, Phys. Rev. D {\bf 106}, 043010 (2022).
   
\bibitem{KS24}
   A. Kumar and H. Sotani, Phys. Rev. D 110, 063001 (2024).

\bibitem{KS25}
   A. Kumar and H. Sotani, Phys. Rev. D {\bf 111}, 123028 (2025).

\bibitem{Ellis18a}  
   J. Ellis, A. Hektor, G. H\"{u}tsi, K. Kannike, L. Marzola, M. Raidal, V. Vaskonen, Phys. Lett. B {\bf 781}, 607 (2018).

\bibitem{Flores24}
   C. V. Flores, C. H. Lenzi, M. Dutra, O. Louren\c{c}o, and J.~D.~V. Arba\~{n}il, Phys. Rev. D {\bf 109}, 083021 (2024).


\bibitem{Comer99}
   G. Comer, D. Langlois, and L. M. Lin, Phys. Rev. D {\bf 60}, 104025 (1999).
   
\bibitem{Comer02}
   G. L. Comer, Found. Phys. {\bf 32}, 1903 (2002).

\bibitem{AC02}
   N. Andersson, G. L. Comer, and D. Langlois, Phys. Rev. D {\bf 66}, 104002 (2002).
   
\bibitem{Comer03}
   G. L. Comer and R. Joynt, Phys. Rev. D {\bf 68}, 023002 (2003).
   
\bibitem{Comer04}
   G. Comer, Phys. Rev. D {\bf 69}, 123009 (2004).

\bibitem{SK25b}
   H. Sotani and A. Kumar, Eur. Phys. J. C,  {\bf 85}, 1438 (2025).

\bibitem{Goldman13}
   I. Goldman, R. Mohapatra, S. Nussinov, D. Rosenbaum, and V. Teplitz, Phys. Lett. B {\bf 725}, 200 (2013).

\bibitem{Sagun23}
   V. Sagun, E. Giangrandi, T. Dietrich, O. Ivanytskyi, R. Negreiros, and C. Provid\^{e}ncia, Astrophys. J. {\bf 958}, 49 (2023).

\bibitem{DD2}  
   S. Typel, Phys. Rev. C {\bf 89}, 064321 (2014).



\bibitem{QMC4}  
   M. G. Alford, L. Brodie, A. Haber, and I. Tews, Phys. Rev. C {\bf 106}, 055804 (2022).


\bibitem{SLy4}
   F. Douchin and P. Haensel, Astron. Astrophys. 380, {\bf 151} (2001).

\bibitem{SKa}  
   H. S. K\"{o}hler, Nucl. Phys. {\bf A258}, 301 (1976).

\bibitem{SkI3}  
   P.-G. Reinhard and H. Flocard, Nucl. Phys. {\bf A584}, 467 (1995).
   

\bibitem{Togashi17}  
   H. Togashi, K. Nakazato, Y. Takehara, S. Yamamuro, H. Suzuki, and M. Takano, Nucl. Phys. A {\bf 961}, 78 (2017).

\bibitem{QHC21}
   T. Kojo, G. Baym, and T. Hatsuda, The Astrophys. J. {\bf 934}, 46 (2022).

   
\bibitem{SIOO14} 
   H. Sotani, K. Iida, K. Oyamatsu, and A. Ohnishi, Prog. Theor. Exp. Phys. {\bf 2014}, 051E01 (2014).

\bibitem{Shlomo06} 
   S. Shlomo, V. M. Kolomietz, and G. Col\`{o}, Eur. Phys. J. A 30, 23 (2006).

\bibitem{Garg18} 
   U. Garg and G. Col\`{o}, Prog. Part. Nucl. Phys. 101, 55 (2018).

\bibitem{Vinas14}
  X. Vi\~{n}as, M. Centelles, X. Roca-Maza, and M. Warda, Eur. Phys. J. A {\bf 50}, 27 (2014).
  
\bibitem{Li19}
   B. A. Li, P. G. Krastev, D. H. Wen, and N. B. Zhang, Eur. Phys. J. A 55, 117 (2019).


\bibitem{Foot91}
   R. Foot, H. Lew, and R. R. Volkas, Phys. Lett. B {\bf 272}, 67 (1991)

\bibitem{KS25b}
   A. Kumar and H. Sotani, Phys. Rev. D 112, 063030 (2025).

\bibitem{Grippa}  
   F. Grippa, G. Lambiase, T. K. Poddar, Universe {\bf 11}, 74 (2025).
   
   
\bibitem{Jungman96}  
   G. Jungman, M. Kamionkowski, and K. Griest, Phys. Rep. {267}, 195 (1996).

\bibitem{Salas21}  
   P. F. de Salas and A. Widmark, Rep. Prog. Phys. {\bf 84}, 104901 (2021).

\bibitem{Ellis18}  
   J. Ellis, G. H\"{u}tsi, K. Kannike, L. Marzola, M. Raidal, and V. Vaskonen, Phys. Rev. D {\bf 97}, 123007 (2018).

\bibitem{Capela13} 
   F. Capela, M. Pshirkov, and P. Tinyakov, Phys. Rev. D {\bf 87}, 023507 (2013).




\bibitem{YK97}
   S. Yoshida and Y. Kojima, Mon. Not. R. Astron. Soc. {\bf 289}, 117 (1997).

\bibitem{Sotani22}
   H. Sotani, Eur. Phys. J. C {\bf 82}, 477 (2022).










\end{thebibliography}

\end{document}